\documentstyle[11pt]{article}

\begin{document}

\title{Intermittency in turbulence}

\author{F. Hayot and C. Jayaprakash \\ Department of Physics, \\
The Ohio State University, \\ Columbus, Ohio 43210}

\maketitle
\begin{abstract}

We derive from the Navier-Stokes equation an exact equation satisfied by
the dissipation rate correlation function,  
$<\epsilon(\vec{x}+\vec{r},t+\tau)\epsilon(
\vec{x},t)>$, which we study in the equal time limit. We exploit its
mathematical similarity to the corresponding   equation derived from
the 1-dimensional stochastic  Burgers equation to show that the main
intermittency exponents are
$\mu_1=2-\zeta_6$ and $\mu_2=2\tilde{z_4}-\zeta_4$, where 
the $\zeta$'s are exponents of
velocity structure functions and $\tilde{z_4}$ is a dynamical exponent 
characterizing the $4$th order structure function. We  
discuss the role of sweeping and Galilean invariance in determining the
intermittency exponents.

\end{abstract}


\section{Introduction}

 The correlations  of
the local energy  dissipation rate per unit mass, $\epsilon(\vec x,t)$, 
whose behavior in the inertial range 
can be written  as
\begin{equation}
<\epsilon(\vec x+\vec r) \epsilon(\vec x)> \sim ~ <\epsilon>^2~(r/L)^{-\mu} ~~~.
\end{equation}
play an important role in the study of high Reynolds number 
turbulence\cite{frisch,my}. The reason is that the algebraic decay of the
correlations  characterized
by a positive value of $\mu$ signals 
intermittent behavior. In Eqn. (1)  $<\epsilon>$ is the mean value of the
dissipation rate per unit mass,  $r$ belongs to the
inertial range, and $L$ is the integral scale or system size. We have assumed
that the turbulence is homogeneous and isotropic.
When $\mu=0$ there are no fluctuations at large distances 
and scaling holds. A positive
intermittency exponent 
$\mu$ signals the breakdown of Kolmogorov scaling and, experimentally this 
is indeed the case; there now appears to be agreement\cite{sreeni} on a value of 
$\mu=0.25 \pm 0.05$. 

The standard description of the statistical properties of turbulence 
in terms of velocities leads to the introduction of (longitudinal)
structure functions, 
$S_p(r)$, whose behavior in the inertial range is given by
$$ S_p(r) ~\equiv~ <~[(u(\vec{r}) - u(\vec{0})] \cdot \hat{r}]~^p> ~~ \sim ~~
r^{\zeta_p} ~~~.$$
 A scaling argument (of which a more
sophisticated version is the refined similarity hypothesis of Kolmogorov and
Obukhov\cite{my}) has been employed to relate the exponent $\mu$
characterizing dissipation to the exponent,
$\zeta_6$, which describes  the behavior of the $6$th order structure function. 
In the naive scaling argument,  since $\epsilon$ has
dimensions
of $velocity^3/length$, one expects that $\epsilon$, when it occurs in
correlation functions,  behaves as 
$(\Delta u)^3/r$, where $\Delta u(r) \equiv [\vec{u}(\vec x +\vec
r)-\vec{u}(\vec x)]
\cdot
\hat{r}$ is the longitudinal velocity difference. Therefore, the dissipation 
correlation in (1) can be expected to decay as
$S_{6}(r)/r^2$.  This reasoning leads to the well-known
identification\cite{frisch}
\begin{equation}
\mu = 2-{\zeta}_6   ~~~~.
\end{equation}
Experimental measurements yield $\zeta_6 \approx 1.8$ (its scaling
value is $2$); the value of $\mu$ which is then 
obtained from Eqn. (2) is consistent with
the experimental result of $\mu=0.25\pm 0.05$ quoted above. Because
of this agreement many discussions concerning the intermittency exponent have
been limited to a discussion of the non-scaling behavior of $\zeta_6$,
although other expressions for $\mu$ have been proposed. In particular, 
the relation
\begin{equation}
\label{nelkineq}
\mu = 2\zeta_{2} - \zeta_{4}
\end{equation}
 has been proposed.\cite{nelkin,fsn} It leads to a value for $\mu$ of
$\mu \simeq 0.1$ using
data in the literature.\cite{frisch131} For $r << L$ , the
larger value of $\mu$ would dominate.

Indeed one should generally expect that in an appropriate 
equation satisfied by  the 
dissipation correlation  different
terms will lead to different possible values for $\mu$
from which the dominant behavior in the inertial
range can be extracted.  As far as we know a systematic treatment
of this question does not exist. The aim of this paper is to provide one.
We present an {\em exact} equation satisfied by the dissipation correlations,
analyze it, and in particular, show how the two aforementioned values
of $\mu$ arise. We should stress here that the only way we have found to
derive the equations for the (experimentally measured) equal
time, spatial correlations is to start from correlations in both space and time,
and then impose the equal time limit.\cite{fn1} 

We will treat both the Navier-Stokes equation and the one-dimensional
stochastic Burgers equation.\cite{burgers} The mathematical structure of 
equations derived from the latter, while simpler algebraically, is similar to
that encountered for the Navier-Stokes equation, as we have pointed out
previously.\cite{hj1,hj2,hj3,hj4} The stochastic Burgers equation thus 
provides a
gateway to understanding issues of fully developed turbulence and our results
confirm the usefulness of this approach. It enables one to go beyond generalized
scaling arguments which essentially embody the correctness (up to intermittent
effects) of the original Kolmogorov argument.

In Section $2$ of the paper we present the 
exact equations for dissipation correlations, both for the Navier-Stokes
equation and the stochastic 
Burgers equation. The details of the straightforward derivation
are relegated to  Appendix A.
Our analysis discussed in Sec. $3$ shows that there are two main
contributions
in the inertial range: the dominant one characterized by the  exponent
$\mu=2-\zeta_6$, and a second which  depends on the {\em dynamical} behavior
of the $4$th order structure function. The mathematical 
similarity of the 
analysis for the Burgers problem and that for the Navier-Stokes equation is
discussed. In section \ref{tech} we turn 
to a discussion of some technical points, in
particular, the role played by Galilean invariance of the two fundamental
equations, and its relationship to ``sweeping" effects. A summary of our results
and conclusions follow. Technical points are taken up in Appendices B, C
and D.

\section{The fundamental equations and their derivation}

In addition to the Navier-Stokes equation we will also investigate, as
mentioned in the Introduction,  the
one-dimensional, stochastic Burgers equation,\cite{burgers} describing a
compressible fluid, given by
\begin{equation}
\label{burgers}
\partial u/\partial t + u\partial u/\partial x = \nu
\partial^2 u /\partial x^2~ +~\eta(x,t)
\end{equation}
where $\nu$ is the kinematic viscosity, and $\eta$ a long-ranged,
Gaussian random noise
with zero mean and correlations in $k$ space given by
\begin{equation}
\label{noise}
<\hat{\eta}(k,t)\hat{\eta}(k',t') = 2 D(k)\delta_{k+k',0}\delta(t-t')
\end{equation}
where $D(k)=D_0|k|^{\beta}$. The exponent  $\beta$ determines the range of
the forcing and we focus on $\beta \in (0,-1]$. We have shown
previously \cite{hj1,hj2} that for these values of $\beta$ the system exhibits a
rich multifractal structure, going from a near scaling regime for
$\beta$ close to zero to an extreme multicritical regime  characterized by the
presence of strong shocks, with $\zeta_n
\rightarrow 1 $ for $n \geq 3$ when $\beta\rightarrow  -1$.

We will derive from the stochastic Burgers equation and analogously in the
Navier-Stokes case, an equation satisfied by the  correlations of the local
dissipation rate. We emphasize that it is convenient to derive  an equation for
the time-dependent quantity
\begin{equation}
\tilde{C}(r,\tau) = <\epsilon(x+r,t+\tau)\epsilon(x,t)>
\end{equation}
where the local dissipation rate is defined by
\begin{equation}
\epsilon ~=~ \frac{\nu}{2} (\partial_i u_j +\partial_j u_i)(\partial_i u_j
+\partial_j u_i)
\end{equation}
in the Navier-Stokes case and by $\epsilon=\nu(\partial u/\partial x)^2$ in the
one-dimensional Burgers problem. We have employed the notation 
$\partial_i \equiv \partial/\partial x_i$.  The equal time correlation function
can be obtained by taking the limit
$\tau
\rightarrow 0$.

Denote the space-time coordinates of two points by $\vec x$
and
$\vec x' =\vec x +
\vec r$ and $t$ and $t'=t+\tau$ and the various fields by primed and unprimed
variables, for example $\epsilon' \equiv  \epsilon(\vec x',t')$, $\epsilon
\equiv \epsilon(\vec x , t)$ and similarly for the velocities and other fields.
We consider the Navier-Stokes equation driven by a random force  
$\hat{\vec{\eta}}(\vec k,t)$ concentrated around a small wavevector; this
enables us to define  average values most simply.  It is straightforward to
derive (see  Appendix A) the
following equation describing the spatio-temporal behavior  of the correlations
of the dissipation rate:
\begin{eqnarray}
\label{nsbasic}
<\epsilon'\epsilon> &=&  \frac{1}{4} \partial_{\tau}<\epsilon
\vec{u'}\cdot \vec{u'}-\epsilon'\vec{u}\cdot\vec{u}>~
 +~\frac{1}{4}\partial_{r_j} <\epsilon \vec{u'}\cdot \vec{u'}u'_j-
\epsilon' \vec{u}\cdot \vec{u}u_j>  \nonumber  \\
&&+\frac{\nu}{4}{\nabla_r}^2<\epsilon'{\vec{u}}\cdot
\vec{u}+\epsilon\vec{u'}\cdot\vec{u'}>
+\frac{\nu}{2}\nabla_{r_i}\nabla_{r_j}<\epsilon'u_{i}u_{j}
+\epsilon u'_{i}u'_{j}> \nonumber \\
&& +\frac{1}{2}<\epsilon'\vec{u}\cdot \vec{\eta}+
\epsilon\vec{u'}\cdot \vec{\eta'}>
+\frac{1}{2}\partial_{r_i}<\epsilon' pu_i-\epsilon p' u'_i>
\end{eqnarray}
where $p$ is the pressure divided by the
mass density. This exact equation is one of the fundamental equations in this
paper.\\

Similarly one can derive (see Appendix A) the corresponding equation for
$\epsilon(x,t)$ in the $1d$
stochastic Burgers equation. We will denote the two space-time points by
$x_1=x+\frac{1}{2}r$, $t_1=t+\frac{1}{2}\tau$, $x_2=x-\frac{1}{2}r$,and
$t_2=t-\frac{1}{2}\tau$ respectively and use the notation $u_1 \equiv
u(x_1,t_1)$, etc., for the various fields. The equation satisfied by
$\tilde {C}(r,\tau) \equiv <\epsilon_1\epsilon_2> is
\equiv <\epsilon(x_1,t_1)\epsilon(x_2,t_2)>$
\begin{eqnarray}
\label{burgbasic1}
 <\epsilon_1\epsilon_2> ~ &=& ~
\frac{1}{4} {\partial}_{\tau}<\epsilon_1 {u_2}^2 - \epsilon_2 {u_1}^2>
+\frac{1}{6} {\partial}_{r}<\epsilon_1{u_2}^3 - \epsilon_2 {u_1}^3> \nonumber
\\
&& +\frac{\nu}{4} \partial_r^2 <\epsilon_1 {u_2}^2 + \epsilon_2
{u_1}^2> \nonumber \\
&& + \frac{1}{2} <\epsilon_1 u_2 \eta_2 + \epsilon_2 u_1
\eta_1> ~~.
\end{eqnarray}
\\

 We draw attention to
the close  mathematical similarity of the two equations, in particular, to the
first two terms in each equation, schematically $\partial <\epsilon
u^2>/\partial \tau$ and $\partial <\epsilon
u^3>/\partial r$,  which yield  the dominant contributions to
the  behavior of the dissipation correlation. The
second line in each equation contains terms proportional to
$\nu$ and these do not contribute in the large Reynolds number limit. We
emphasize that the
$\nu-$dependent terms that are multiplied by singular operators have to be
handled carefully to extract the innocuous terms in the second line (see
Appendix A). The
difference in the last lines is evidently due to the pressure term in the
Navier-Stokes equation. We hasten to add that physically the nature of
turbulence is {\em not} the same in the two systems. For example, vortex
stretching is important in the three-dimensional  problem. In addition,
the energy flux through a wavenumber
$k_0$, $\Pi(k_0)$, is proportional to
$(D_0/L)k_0^{1+\beta}$ in the stochastic Burgers problem while it is a constant
in the Navier-Stokes case. Nevertheless, the mathematical similarity permits a
parallel analytical investigation.
Thus the  main
features of Eqn. (\ref{nsbasic}), namely the expressions for the dominant
intermittency exponents quoted in  the Introduction,
can be extracted by learning to analyze the
equation in 
the  one-dimensional stochastic Burgers equation and employing a similar
strategy in the Navier-Stokes case. These equations for the
dissipation rate correlations depend crucially on including time derivatives and
the variation of the temporal fluctuations contribute to the potential
intermittent behavior of
$<\epsilon
\epsilon'>$. \\

\section{Analysis of the fundamental equation in the $1d$ Burgers case}

We will begin with the mathematically simpler one-dimensional case; in the rest
of the paper we restrict our attention  to the limit 
$\tau \rightarrow 0$.  In this limit the last term in Eqn.
(\ref{burgbasic1}) which depends on the random force is equal to the product
$<\epsilon_1><\epsilon_2>$, apart from terms that vanish in the limit $\nu
\rightarrow 0$. This is shown  most easily in $k$ space,
given the correlation of the random force in Eqn. (\ref{noise}), using
Novikov's result for Gaussian random processes\cite{novikov}
and the
result $<\epsilon> = \frac{1}{L^2}\sum_{k} D(k)$. This leads
in the $\tau \rightarrow 0$ limit to
\begin{eqnarray}
\label{burgbasic2}
 C(r) \equiv  <\epsilon_1 \epsilon_2> - <\epsilon>^2  &= &
\frac{\nu}{4}[~\partial_r^2 <\epsilon_2 {u_1}^2 +
\epsilon_1 {u_2}^2> +\partial_r S_2 (\frac{1}{6} \partial_r^2 S_3 - \nu
\partial_r^3 S_2)~]
 \nonumber \\
 && +\frac{1}{6} {\partial}_{r}<\epsilon_2 {u_1}^3 -
\epsilon_1 {u_2}^3>
 +\frac{1}{4} {\partial}_{\tau}<\epsilon_2 {u_1}^2 - \epsilon_1 {u_2}^2>~~.
\end{eqnarray}
All the terms on the right-hand side are evaluated in the
$\tau \rightarrow 0$ limit.

\subsection{Outline of the argument and main results}
\label{out}

We analyze the fundamental equation in the Burgers case, Eqn.
(\ref{burgbasic2}), for $r$ in the inertial range and extract the dominant
terms. In the zero-viscosity limit it is easy to see that the first line in Eqn.
(\ref{burgbasic2}) does not contribute: clearly, terms that do not
depend explicitly on
$\epsilon$  are finite  and
hence, are negligible as $\nu \rightarrow 0$. Note that in contrast to the case
of fully developed turbulence where $<\epsilon>$ is a constant, in the stochastic
Burgers equation one has 
$<\epsilon> \propto \delta^{-1-\beta}$ 
where $\delta$ is an inner cutoff scale (shock thickness) and since $\nu \propto
\delta^{1-\beta/3}$ the other terms proportional to $\nu$ vanish for $\beta <
0$.\cite{burgnu}

The crux of the argument for extracting the possible values of $\mu$ depends on
the observations that the dominant behavior of
 $<\epsilon_1 {u_2}^3 -
\epsilon_2 {u_1}^3>$ is determined by $\partial S_6(r,\tau)/\partial
r=\partial/\partial \tau <u_1-u_2)^6>$ and that of
$<\epsilon_1 {u_2}^2 - \epsilon_2 {u_1}^2>$ primarily by $\partial S_4
/\partial \tau$ (we discuss this later) and thus the dominant singular behavior of the
dissipation correlation contains
$$ \frac{\partial^2S_6}{\partial r^2} ~~~~~\mbox{and}~~~~~
\left(\frac{\partial^2S_4}{\partial \tau^2} \right )_{\tau=0}~~.$$
Since $S_6(r) \sim r^{\zeta_6}$ in the inertial range
the  first term immediately yields the well-known intermittency exponent
\begin{equation}
\label{mu1}
\mu_1 = 2-\zeta_6
\end{equation}
for the large distance behavior of $C(r)$.

The second term  leads to an intermittency exponent
\begin{equation}
\label{mu2}
\mu_2 = 2\tilde{z_4}-\zeta_4
\end{equation}
where the exponent $2\tilde{z_4}$  determines the dynamical behavior of
the second derivative of the fourth order structure function when $\tau
\rightarrow 0$. If one assumes {\em naive} dynamic scaling\cite{lpp} 
\begin{equation}
S_4(r,\tau) ~\approx~ |r|^{\zeta_4}~{\cal{S}}_4(|r|/|\tau|^{z_4})
\end{equation}
one has 
$\tilde{z_4} = z_4$ and thus 
\begin{equation}
\label{mu2scaling}
\mu_2 = 2z_4-\zeta_4   ~~~~.
\end{equation}
 We will  show later that
similar results obtain in the Navier-Stokes case also. Accepting this for now 
and using the fact that in the scaling limit,  there is only one 
dynamical exponent $z=z_4$ which has the same numerical value 
as $\zeta_2=2/3$  in the Navier-Stokes case\cite{frisch}, 
the second intermittency exponent can be written as 
$$\mu_2 = 2\zeta_2-\zeta_4 $$
which is  Eqn. (\ref{nelkineq}).
This is precisely the result obtained in reference \cite{nelkin} 
using scaling arguments
for static expressions. Our
analysis shows that the expression in Eqn. (\ref{nelkineq}) is an approximation,
in the scaling limit, of what is actually a dynamical result.  
In the Navier-Stokes case if Kolmogorov scaling
holds $\mu_1=\mu_2=0$. 
Returning to the
stochastic Burgers equation, for small
$\beta$ where all the relevant structure functions scale
$\mu_1=\mu_2=2+2\beta$; 
 in the limit $\beta \rightarrow -1$ where $\zeta_2=2/3$,  the Kolmogorov value
in $3d$ turbulence,   the scaling values of both $\mu_1$ and
$\mu_2$ vanish as in the Kolmogorov case.  Note, however, that $\mu_1 \ge 1$ in
the range
$\beta
\in (0,1]$.  As $\beta \rightarrow -1$ and multiscaling occurs for $p \ge 4$ ,
$\mu_1= 2-\zeta_6 $ approaches unity from above and dominates the behavior 
of $C(r)$.\cite{burgmu2}  We draw attention to the fact that for the
dynamical exponent in Eqn. (\ref{mu2})
we have {\em not} used the value of the  kinematic exponent
$z=1$ that arises from sweeping but the intrinsic  dynamical exponent. In Sec.
\ref{tech} we provide a more detailed justification for this. \\

\subsection{Contribution of the  $<\epsilon_1u_2^3-\epsilon_2u_1^3>$ term}

We will now argue that $<\epsilon_1u_2^3-\epsilon_2u_1^3>$ behaves as
$\partial_r S_6$ for $\tau=0$ as claimed above. In an earlier paper\cite{hj3} we
have presented the equations satisfied by the (low-order) equal-time structure
functions; in particular, the sixth-order structure function, 
$S_6$,  satisfies the equation
\begin{eqnarray}
\label{burgds6dr}
\frac{\partial S_6(r,\tau=0)}{\partial r} ~&=&~ 3\nu{\partial_r}^2S_5(r)
~-~30<(\epsilon_1+\epsilon_2)(u_1-u_2)^3>~+~
\frac{15}{2}<(u_1-u_2)^4(\eta_1-\eta_2)>
\nonumber \\
&=& 3\nu{\partial_r}^2S_5(r) ~-~30[<(\epsilon_1+\epsilon_2)
(u_1-u_2)^3>-<\epsilon_1+\epsilon_2>S_3] ~+~ 5 S_3(r) dS_3/dr
 \nonumber \\
\end{eqnarray}
In obtaining the second equality, the noise term has been  simplified using
Novikov's theorem and the von-Karman-Howarth relation.
The contribution  $S_3~dS_3/dr$ that arises from the noise term can clearly be
identified as the scaling contribution to
$S_6$  which
yields  for the exponent describing the inertial-range behavior 
$\zeta_6 = 2\zeta_3 =-2\beta$. The $\nu-$dependent contribution vanishes as $\nu
\rightarrow 0$ since it is multiplied by a term which is finite in the inertial
range and thus we have,
generically, the result that terms of the form
$<\epsilon_1u_2^3>$ are  proportional to $dS_6/dr$. 
This
immediately yields the result
$\mu_1=2-\zeta_6$ as stated earlier. We emphasize the feature
that the only term that can possibly lead to non-scaling behavior of $S_6$ is 
the term which is a product of the dissipation rate and velocities. This
result, derived
from the exact equation for $S_6$, agrees with
the field-theoretic point of view in which the non-scaling behavior arises 
from  short-distance singularities which are determined by the limit $\nu
\rightarrow 0$ and the existence of an energy cascade.

\subsection{Contribution of the  $<\epsilon_1u_2^2-\epsilon_2u_1^2>$ term}
\label{contrib}

In contrast to the contribution of
$\partial_r <\epsilon_1u_2^3-\epsilon_2u_1^3>$ to the dissipation correlation
function, the term $\partial_{\tau}<\epsilon_1u^2_2-\epsilon_2u_1^2>|_{\tau=0}$
is an intrinsically dynamical object.  In order to study it we consider the 
temporal evolution of
$S_4(r,\tau) \equiv <(u(x_1,t_1)-u(x_2,t_2))^4>$ where $\tau=t_1-t_2$ and
$r=x_1-x_2$. It can
be shown to satisfy the equation
\begin{eqnarray}
\label{ds4dtau}
\partial S_4/\partial {\tau}~& = &~
-\frac{1}{2}{\partial}_r<(u_2-u_1)^4(u_2+u_1)>~+2<(u_1-u_2)^3(\eta_1+\eta_2)>
 \nonumber \\
&& -6<(\epsilon_1-\epsilon_2) (u_2-u_1)^2>    ~~~~.
\end{eqnarray}
The derivation of the general equation satisfied by the time evolution of the
generating function for 
$S_q(r,\tau)$ is given in Appendix B. The above equation follows directly from 
it as a special case. This equation contains a term of the generic form
$<\epsilon u^2>$ whose derivative with respect to $\tau$ occurs in the basic
equation for $<\epsilon_1 \epsilon_2>$. The analysis of this equation is somewhat more 
subtle; however, once
again we expect that the non-scaling behavior of the left-hand side 
is determined
by correlations that involve products of the dissipation rate and velocities,
i.e., the last term on the right-hand side.  As in the previous subsection we
expect the noise term to yield a scaling result: to see this we first note that 
in the limit $\tau
\rightarrow 0$ the noise term yields a discontinuity since it gives
a different contribution depending on whether 
$\tau
\rightarrow 0^+$ or $\tau \rightarrow 0^-$. A careful analysis yields,
 for $\tau >0$, a contribution 
$$-\frac{1}{2} S_2(r) \frac{dS_3(r)}{dr} ~~$$ which dominates the behavior of
$\partial_{\tau}S_4|_{\tau=0}$. In the
range of $\beta$ we are interested in both $S_2$ and $S_3$ exhibit scaling 
behavior. \cite{fn2}  Thus, the noise term again yields scaling behavior for the dynamics
of the first derivative in the
$\tau \rightarrow 0$ limit. The first term in Eqn. (\ref{ds4dtau}) ensures 
Galilean invariance of
the equation and contains only sweeping effects as will be discussed below.
Therefore, the behavior of terms of the
form $[\partial_{\tau} \epsilon_1u_2^2]_{\tau=0}$ in the equation
for 
$<\epsilon_1 \epsilon_2>$  will reflect the non-trivial
singularity in
$\partial^2_{\tau} S_4|_{\tau=0}$, which  then yields the second
intermittency exponent, expression (12), namely 
$\mu_2=2\tilde{z_4}-\zeta_4$. \\

\subsection{Technical remarks}
\label{tech}

In this section we discuss some of the technical points that were not addressed
in the earlier sections. There are two related difficulties: the first is the 
more obvious one;  while our  equation for
$<\epsilon_1\epsilon_2>$ (see Eqn. (\ref{burgbasic2})) 
involves velocities, our equations for $\partial_r
S_6$ and $\partial_{\tau} S_4$ involve velocity differences and the precise
combinations of the terms do not match exactly. Second, 
if we carry out a straightforward  analysis of the term  $\partial_{\tau}
<\epsilon_1u_2^2-\epsilon_2u_2^2>$  we obtain terms of the form  
$\partial_r^2
<(u_1-u_2)^4(u_1^2+u_2^2)>$. A naive factorization of the expectation value 
leads to an exponent $2-\zeta_4$ and not the value of $\mu_2$ 
claimed earlier. In
particular, this corresponds to the kinematic 
value $z_4=1$ in the expression for
$\mu_2$ (Eqn. (\ref{mu2scaling})) and not 
to the dynamical value of $z_4$, which
in the scaling limit is numerically equal to $\zeta_2$. While we
have not been able to provide a mathematically complete analysis of all 
these points  we present below convincing arguments in support of our results 
based on general considerations. 
\\

The first observation we make is that while $<\epsilon_1\epsilon_2>$ is
manifestly Galilean invariant the invariance of the right hand side of Eqn.
(\ref{burgbasic1}) is not evident; it can be checked explicitly (see Appendix
C). Therefore, we expect  a (Galilean-invariant) operator product 
expansion analysis of the terms of the form
$\partial_{\tau} <\epsilon_1u_2^2-\epsilon_2u_1^2>$ 
to pick out precisely the terms that we have used in our analysis  
since the terms will involve Galilean invariant
combinations of operators such as, for example,  velocity differences.
Physically, the potentially troublesome terms, such as $2-\zeta_4$ mentioned
above, arise from the effects of sweeping. For example, we have discussed 
in a previous paper\cite{hj4} how 
behavior such as 
$$<(u_1-u_2)^4(u_1^2+u_2^2)> \sim <(u_1-u_2)^4><(u_1^2+u_2^2)>$$
arises. 
For {\em nonzero} values of $\tau$  such a Galilean non-invariant term is known
to contain effects of sweeping.  That our formalism  gives rise to
such effects is not surprising since we have adopted
an Eulerian point of view.  As pointed out by
Tennekes\cite{tennekes} large-scale energy containing structures advect
inertial-range information past an Eulerian observer. This allows one to note
that the dominant term in $<(u_1-u_2)^4(u_1^2+u_2^2)>$ is proportional to
$<u^2> <(u_1-u_2)^4>$ where we have exploited the independence of the
large-scale flow characteristics encapsulated in $<u^2>$ and the information
about the inertial range contained in $<u_1-u_2)^4>$. Good numerical support
for the correctness of such an argument has been presented in ref.
\cite{hj4}. 
In other words, in 
the Eulerian picture the dynamical behavior of the
structure functions at finite values of the time difference
$\tau$ is dominated by sweeping effects which are characterized by 
an effective
dynamical exponent  of $z=1$. (The second intermittency exponent, $\mu_2$,
shown earlier to be $2\tilde{z}_4-\zeta_4$ would then reduce precisely 
to $2-\zeta_4$ as discussed
in the previous paragraph. )
However here we are dealing with {\em the limit} $\tau
\rightarrow 0$.  Physically, {\em equal time correlations do not 
show sweeping effects}. 
Therefore, only the intrinsic, i.e., non-sweeping, dynamical behavior 
contributes to $<\epsilon_1 \epsilon_2>$. In support of this, 
we also point out that we have shown previously 
\cite{hj4} that  intrinsic dynamical fluctuations determine the
behavior of
$\partial_{\tau} S_2|_{\tau=0}$ and yield the exponent $z_2=1+\beta/3$, while
the correlation $S_4(r=0, \tau)$, where one observes fluctuations 
at a given point as a
function of time picks up the sweeping contribution of exponent $z=1$.\\
Experimentally, given the validity of Taylor's frozen turbulence hypothesis,
especially in situations with a large, externally imposed flow, one can 
determine equal time correlations by making one point measurements as a
function of time. In fact, the transition from spatial to temporal variables
depends on the kinematic exponent $z=1$. Once,  the equal time 
dissipation rate correlations are thus 
determined, albeit approximately, their behavior is given by our analysis.\\

One other term in $C(r)$ not considered thus far 
is the $\tau \rightarrow
0$ limit of 
$$ \frac{\partial}{\partial \tau}<(u_1-u_2)^3(\eta_1+\eta_2)>$$. This term
results from taking the $\partial/\partial \tau$ of Eqn.
(\ref{ds4dtau}). The term 
$<(u_1-u_2)^3(\eta_1+\eta_2)>$ can be shown to be related to a fourth-order 
response function and a naive argument yields inertial-range behavior
characterized by an exponent that is the scaling value of $\mu_2$.

Thus, the behavior of $C(r)$ in the $\tau \rightarrow 0$ limit is  determined
by appropriate Galilean invariant combinations of spatial and temporal
derivatives of structure functions, such as 
$\frac{\partial^2S_6}{\partial r^2}$
or
$\partial^2_{\tau}<(u_1-u_2)^4>+\partial_{\tau}\partial_r
<(u_1-u_2)^4(u_1+u_2)>+(1/4)\partial^2_r <(u_1-u_2)^4(u_1+u_2)^2>$, and they
give rise to the dominant exponents discussed above. Of course, we have not provided a
first-principles determination of the dynamical exponent $z_4$, or for that
matter, of $\zeta_6$. We aimed  only at establishing relations for the possible
value of the intermittency exponent $\mu$.

\section{Analysis of the Fundamental equation for the Navier-Stokes case}

We will follow in the 
three-dimensional Navier-Stokes case the strategy outlined earlier for the 
stochastic Burgers equation. The calculations  are
much more involved to carry out in the same detail. We will point out the
mathematical correspondences between terms in  the 
Navier-Stokes and
the Burgers cases: we obtain the vectorial or tensorial  generalizations
of the scalar terms in the $1d$ Burgers problem that are allowed by
rotational invariance. For example, the analysis of 
the crucial $<\epsilon (\delta
u)^3>$ term in Eqn. (\ref{burgbasic2}) that  gave rise to 
$\mu=2-\zeta_6$ was
based on the equation for $\partial_r S_6$. Thus we are led to  consider the
corresponding equation for an appropriate sixth-order structure function in the
Navier-Stokes case. We use the notation

$$\delta \vec{u}~ \equiv~ \vec{u}(\vec x', t') ~-~ \vec{u}(\vec x,t)
~\mbox{and}~ \delta \vec{\eta}~\equiv~ \vec{\eta}(\vec x',t')~-~ 
\vec{\eta}(\vec x,t)~~.$$
for velocity and noise differences. We have derived the following equation
\begin{eqnarray}
\label{nsds6dr}
&&\frac{\partial~ }{\partial r_j} < (\delta \vec u \cdot \delta \vec u)^2
 \delta u_j \delta u_k > ~=~
\frac{\nu}{2} \nabla_r^2 < [\delta \vec u \cdot \delta \vec u]^2 \delta u_k >
\nonumber \\
&&- 4 <\delta \vec u \cdot \delta \vec u ~\delta u_k
[\hat{\epsilon}+ \hat{\epsilon}']> -6 <\delta \vec u \cdot \delta \vec u 
~\delta u_i [\hat{\epsilon}_{ki}+\hat{\epsilon}'_{ki}] > -2
<\delta u_i \delta  u _j ~\delta u_k [\hat{\epsilon}_{ij}+\hat{\epsilon}'_{ij}] > 
\nonumber \\ && < [\delta \vec u \cdot \delta \vec u]^2 \delta \eta_k + 4 \delta u_k
\delta \vec u \cdot \delta \vec u \delta \vec u \cdot \delta \vec \eta>
\nonumber \\
&&-2 <[\delta \vec u \cdot \delta \vec
u]^2 \frac{\partial (p+p')}{\partial r_k}>-4
<\delta \vec u \cdot \delta \vec
u \delta u_k \delta u_i \frac{\partial (p+p')}{\partial r_i}>
\end{eqnarray}
In the preceding we have used the definitions for a (symmetric) tensor with the
dimensions of the energy dissipation rate,
\begin{equation}
\label{disstensor}
\epsilon_{jk} ~=~ \nu \partial_i u_j~ \partial_i u_k ~~.
\end{equation}

The analogy of Eqn. (\ref{nsds6dr}) with the corresponding equation in the
Burgers case (see eqn. (\ref{burgds6dr})) is striking. The first terms in
both equations have the same form and 
vanish in the $\nu \rightarrow 0$ limit. The driving term $<(\delta u )^4
\delta \eta>$ in the Burgers problem generalizes to the two vector terms in the
third line of Eqn.  (\ref{nsds6dr}). The scalar term  $<(\delta u)^3 (\epsilon
+\epsilon')>$, which is crucial for identifying $\mu_1$  in the Burgers
problem,  generalizes in the
Navier-Stokes case to  the three terms in the
second line of Eqn. (\ref{nsds6dr}) and involves the tensor generalization of
$\epsilon$ defined above. It is reasonable to expect  that 
terms on the right-hand side generically of the form
$\epsilon u^3$ which involve the dissipation rate lead to the intermittent 
 behavior of
$\frac{\partial~ }{\partial r_j} < [\delta \vec u \cdot \delta \vec u]^2
 \delta u_j \delta u_k >$ since the dissipation rate contains the singular
behavior of operators in the 
$\nu
\rightarrow 0$ limit. The other difference with the Burgers case 
arises from the pressure terms in the last line of Eqn.  (\ref{nsds6dr}) and we
do not expect it to yield the dominant intermittent behavior of the right-hand
side since it does not depend explicitly on $\epsilon$; we expect it to yield
scaling contributions. 
  Straightforward tensor analysis shows that  the
left hand side of Eqn. (\ref{nsds6dr}) effectively 
includes the contribution of the
sixth-order longitudinal structure function. We do not enter into the issue of
whether longitudinal and transverse structure functions yield the same
exponents; if they do differ as has been claimed \cite{lohse} then our results
should be modified appropriately. The situation is simpler if as is argued by
L'vov {\em et al.}\cite{lpp2} there is a single exponent.

We have also derived the equation for the temporal derivative of an appropriate
fourth-order structure function. The details of the derivation are presented in
Appendix D; the equation is given by 
\begin{eqnarray}
\label{nsds4dtau}
\partial_{\tau} < [\delta \vec u \cdot \delta \vec u]^2  > ~&=&~ -\frac{1}{2}
\partial_{r_j} < (u_j+u'_j)[\delta \vec u \cdot \delta \vec u]^2> 
+2 \langle [\delta \vec u \cdot \delta \vec u]^2 \delta \vec{u} \cdot
(\vec{\eta}+\vec{\eta'}) \rangle   \nonumber \\
&&-2<~(\epsilon-\epsilon') \delta \vec u \cdot \delta \vec u ~> -
4<~(\epsilon_{ik}-\epsilon'_{ik}) \delta u_i \delta u_k ~> -
\nonumber \\
&& + 2\nu \langle \delta \vec u \cdot \delta \vec u \nabla^2 \delta p 
\rangle \nonumber \\
&& ~-~ 4 \langle~ 
\delta \vec u \cdot \delta \vec u \frac{\partial}{\partial r_i}(\delta p \delta
u_i) ~ \rangle
\end{eqnarray}
Again we note the similarity to the structure of the equation in the $1d$
case, Eqn. (\ref{ds4dtau}), which leads to the second intermittency exponent
$\mu_2$. The first term on the right-hand side which ensures the Galilean
invariance of the equation is the vector version of the first term on the
right-hand side of  Eqn. (\ref{ds4dtau}) as is the noise term. The term that
has the generic form $<\delta\epsilon (\delta u)^2>$ which is a scalar in the
Burgers case involves the tensor generalization of $\epsilon$ defined
earlier. It is easy to see that the term proportional to $\nu$ is negligible in
the high Reynolds number limit.  The last term in Eqn. (\ref{nsds4dtau}) which 
depends on the pressure  has no counterpart in the Burgers case; as argued
earlier we do not expect this term to dominate the intermittent
behavior of the left-hand side.  On physical grounds it is reasonable to  expect
that the leading intermittent behavior of
$\partial_{\tau} < [\delta\vec u \cdot\delta \vec u]^2  >$ is determined by
terms that involve the dissipation rate on the right hand side,  of the form
$<\delta\epsilon (\delta u)^2>$, as in the Burgers case; thus  the 
Galilean-invariant contribution of
$\partial_{\tau}<\epsilon
\vec{u'}\cdot \vec{u'}-\epsilon'\vec{u}\cdot\vec{u}>$ to the dissipation
correlations in Eqn. (\ref{nsbasic})  will be given by
$\partial^2 S_4/\partial
\tau^2|_{\tau=0}$. 

We see that the mathematical similarity of the various equations in the Burgers
and Navier-Stokes cases allows us to carry out a similar analysis and deduce the
same relations for the leading intermittency exponents in terms of the
exponents that characterize the static and dynamic behavior of the velocity
structure functions. We have discussed the implications of this already in Sec.
\ref{out}. 

\section{Discussion of Results}

It is remarkable that the exponents which
characterize the power law decay of the (equal time) dissipation rate
correlations  obey  the same relations
for the Burgers equation as for the Navier-Stokes equation. 
The leading exponent $\mu_1=2-\zeta_6$ derives from the presence of
$\partial_r^2 S_6$
in the equation for dissipation fluctuations while the second exponent
$\mu_2$ arises from
the term $\partial_\tau^2 S_4|_{\tau=0}$. Clearly,
the origin of the second exponent (expression (14)) is
dynamical,  and coincides, in the Navier-Stokes case,  with the
static expression in Eqn. (\ref{nelkineq}) (see Ref. \cite{nelkin})
in the scaling limit,  where the only
dynamical exponent $z$ is numerically equal to $\zeta_2$.

We have derived the exact equation satisfied by $C(\vec{r},\tau)$ and
analyzed it to obtain
our results. Recently, L'vov and Procaccia \cite{lp1,lp2} have found general 
fusion rules for
multipoint correlations in the Navier-Stokes problem and used them to show  that
$\mu=2-\zeta_6$; they also discuss the possibility of another scenario (see
Eqn. (15) in Ref. \cite{lp1}) in which the static version of
$\mu_2$, i.e, $\mu_2=2\zeta_2-\zeta_4$ can occur.\cite{nelkin} Our
approach finds these as possible terms from an exact equation and provides a
dynamical expression for the second exponent.
 
In our analysis we have had to disentangle sweeping effects
from the dynamics of internal evolution, both of which are
present in our Eulerian equations.
We did this by  exploiting the Galilean invariance of the internal dynamics,
noting that  
 sweeping  which leads to a kinematic value of $z=1$ yields effects that break
Galilean invariance.  It is clear that for the measurable
quantity we are considering, namely dissipation rate correlations in the inertial
range at equal times, sweeping effects do not matter. As we have shown previously
\cite{hj4} for the Burgers equation, for correlations involving velocities at
different times  sweeping effects can  dominate the dynamic behavior: the
structure functions, for example, 
$S_2$ satisfy a  wave equation with a characteristic velocity
of the order of the rms fluctuations of velocity.

While we have not proved that $\mu_1$ and
$\mu_2$ are the only dominant exponents, especially in the case of the
Navier-Stokes equation, our equations are exact, and the
identifications we have made in 
the 1-dimensional Burgers equation are transparent. The mathematical
similarity of the equations derived from  the Navier-Stokes equation to those in
the Burgers case  gives one
confidence that one has indeed made progress in the derivation of the 
intermittency exponents in the Navier-Stokes
case.
\newpage

\appendix \section{Derivation of Eqns. (8) and (9) }

Starting from the Burgers equation for $u_1=u(x_1,t_1)$ and multiply by $u_1$
to obtain
\begin{equation}
\label{derive1}
u_1 \partial u_1/\partial t_1 ~=~ \nu u_1\partial^2u_1/\partial x_1^2  ~-~
\frac{1}{3} \partial u_1^3/\partial x_1 ~+~ u_1 \eta_1 ~~~.
\end{equation}
The first term on the right-hand side can be rewritten using elementary
manipulations  as
\begin{eqnarray}
\label{derive2}
\nu  u_1\partial^2u_1/\partial x_1^2 ~&=&~ \nu[\frac{\partial (u_1\partial
u_1/\partial x_1)}{\partial x_1} ~-~ (\partial u_1/\partial x_1)^2] \nonumber \\
&=& \frac{\nu}{2}\partial^2 u_1^2/\partial x_1^2 ~-~ \epsilon_1 ~~~~~.
\end{eqnarray}
Note that this is a key step since it allows us to disentangle the terms that
are finite in the $\nu \rightarrow 0$ limit, due to the singular behavior
of the derivative terms, from those that vanish. Using this result and
multiplying Eqn. (\ref{derive1}) by
$\epsilon_2$ yields
\begin{equation}
\label{derive3}
\frac{1}{2}\frac{\partial[\epsilon_2  u_1^2]}{\partial t_1} ~=~
\frac{\nu}{2}\frac{\partial^2
[\epsilon_2 u_1^2]}{\partial x_1^2} ~-~ \epsilon_1\epsilon_2   ~-~
\frac{1}{3} \frac{[\partial \epsilon_2 u_1^3]}{\partial x_1} ~+~ \epsilon_2 u_1
\eta_1
\end{equation}
where we have used the fact that $\epsilon_2$ does not
depend on the coordinates $x_1$ and $t_1$. We compute next the average of the
above equation with respect to the noise ensemble described by Eqn.
(\ref{noise}).  We assume that we have a spatially homogeneous
and temporally steady state which implies that
\begin{equation}
\label{steadyhomog}
\partial <...>/\partial t ~=~ 0 ~~\mbox{and}~~ \partial <...>/\partial x ~=~
0  ~~.
\end{equation}
Using the definitions $x_1=x+r/2$ and $x_2=x-r/2$, rudimentary calculus
yields
\begin{equation}
\label{derivatives}
\partial/\partial x_1 = \frac{1}{2}\partial/\partial x +\partial/\partial r
~\equiv~ \frac{1}{2}\partial_x  + \partial_r
~~\mbox{and}~~ \partial/\partial x_2 = \frac{1}{2}\partial/\partial x
-\partial/\partial r ~\equiv~ \frac{1}{2}\partial_x  - \partial_r
\end{equation}
and similar results for the time variable $t_1$ and $t_2$.
Averaging Eqn. (\ref{derive3}) yields
\begin{equation}
\label{appburg}
\frac{1}{2} \partial_{\tau}<\epsilon_2  u_1^2> ~=~
\frac{\nu}{2} \partial_r^2
<\epsilon_2 u_1^2> ~-~ <\epsilon_1\epsilon_2>   ~-~
\frac{1}{3} \partial_r <\epsilon_2 u_1^3>~+~ <\epsilon_2 u_1
\eta_1>  ~~~.
\end{equation}
Writing a similar equation with the subscripts $1$ and $2$ interchanged and
adding yields the symmetric form displayed in Eqn. (\ref{burgbasic1}).\\

The equation for the dissipation rate correlations in the
Navier-Stokes case Eqn.(\ref{nsbasic}) can be obtained analogously taking
care to keep track of the Cartesian coordinate subscripts and the pressure
term. One finds instead of Eqn. (\ref{appburg}), changing to primed 
variables as is done in the 
main text, the following equation
\begin{eqnarray}
\label{appns}
\frac{1}{2} \partial_{\tau}<\epsilon' \vec{u}\
\cdot \vec{u}> &=&
\frac{\nu}{2}{\nabla_r}^2<\epsilon'\vec{u}\cdot\vec{u}>
+\nu\nabla_{r_i}\nabla_{r_j}<\epsilon' u_{i}u_{j}> \nonumber \\
&& -<\epsilon'\epsilon>
 - \frac{1}{2} \partial_{r_j} <\epsilon' \vec{u}\cdot \vec{u}u_j>  
\nonumber  \\
&& +<\epsilon'\vec{u}\cdot \vec{\eta}>
-\partial_{r_i}<\epsilon' p u_i>
\end{eqnarray} 
The equations for the Burgers and the Navier-Stokes case 
correspond to each other term by term ( if one makes
allowance for the fact that the one deals with one dimensional the other 
with three
dimensional quantities) as is evident from the comparison
of equations (\ref{appburg}) and (\ref{appns}). 
The two additional terms in the Navier-Stokes case
are due to the presence of pressure and the use of the full three 
dimensional definition of $\epsilon$, rather than its expression along
one direction which is the only quantity usually measured \cite{sreeni}.

\section{Derivation of Eqn. (16) }

The general derivation of the equation satisfied by $S_q(r,\tau)$ is most
easily carried out using generating functions. We use  the usual notation
$u_1=u(x_1,t_1)$ and $u_2=u(x_2,t_2)$, $x_1=x+r/2$, etc. Consider the
generating function $\exp a(u_1-u_2)$ and compute its derivative with respect
to $\tau$. Starting from the stochastic  Burgers equation for $u_1$ and $u_2$ 
and using $\partial_{\tau} = (1/2)(\partial_{t_1}-\partial_{t_2})$,
we obtain 
\begin{eqnarray}
\frac{\partial e^{a(u_1-u_2)}}{\partial \tau} ~~&=&~~ \frac{a}{2}
(\partial_{t_1} u_1 + \partial_{t_2} u_2) e^{a(u_1-u_2)} \nonumber \\
&=& \frac{a}{2}e^{a(u_1-u_2)} (\nu \partial_1^2 u_1 -u_1\partial_1 u_1 +
\eta_1~+~ 
\nu \partial_2^2 u_2 -u_2\partial_2 u_2 + \eta_2)  ~~~.
\end{eqnarray}
The following identities that are
easily verified are useful:
\begin{eqnarray}
e^{a(u_1-u_2)}~ u_1 \partial_1 u_1  ~~&=&~~\frac{1}{a^2} \partial_1
[~e^{a(u_1-u_2)} (au_1-1) ~]\nonumber \\
e^{a(u_1-u_2)}~ \partial_1^2 u_1  ~~&=&~~ \frac{1}{a} 
[~\partial_1^2 e^{a(u_1-u_2)}~ -~a^2 e^{a(u_1-u_2)} (\partial_1 u_1)^2 ~] ~~.
\end{eqnarray}
We use these and the corresponding equations for derivatives with respect to
$x_2$ and take average values over the Gaussian noise ensemble. We use Eqn.
(\ref{derivatives}) and  assuming a spatially homogeneous, temporally
steady-state which implies Eqn. (\ref{steadyhomog})
we can evaluate $< \partial e^{a(u_1-u_2)}/\partial \tau>$ and obtain
\begin{eqnarray}
\frac{\partial }{\partial \tau}~\langle e^{a(u_1-u_2)} \rangle ~~&=&~~
-\frac{1}{2}
\partial_r ~\langle -\frac{a^2}{2}\langle  e^{a(u_1-u_2)}
(u_1+u_2) \rangle ~+~ 
\langle e^{a(u_1-u_2)} (\epsilon_1-\epsilon_2) \rangle  \nonumber \\
&+&  \frac{a}{2} \langle e^{a(u_1-u_2)} (\eta_1-\eta_2) \rangle ~~~.
\end{eqnarray}
The coefficient of $a^4$ yields Eqn. (\ref{ds4dtau}) used 
in Sec. \ref{contrib}.

\section{Galilean invariance of Eqn. (9)}

In this Appendix we show the Galilean invariance of the righthand side of 
Eqn. (\ref{burgbasic1}) explicitly. A similar derivation can be carried out in
the Navier-Stokes case. In the one-dimensional problem the Galilean
transformation to a frame moving with a relative velocity $V$ is accomplished
by letting 
$$x ~\rightarrow~ x+Vt ~~~~~\mbox{and}~~~~ t ~\rightarrow ~ t$$
and 
$$\partial/\partial x ~\rightarrow~ \partial/\partial x ~~~~~\mbox{and}~~~~ 
\partial t ~\rightarrow~ \partial/\partial t - V \partial/\partial x ~.$$
Analogous results apply for the relative  coordinates $r=x_1-x_2$ and
$\tau=t_1-t_2$. Making these substitutions in the righthand side of Eqn.
(\ref{burgbasic1}) we find that terms 
proportional to  $V^3$ and $V^2$ vanish as  can be easily
checked. The term of order $V$ is given by 
$$\frac{1}{2} \frac{\partial}{\partial \tau} < \epsilon_1u_2-\epsilon_2u_1>
+\frac{1}{4} \frac{\partial}{\partial r} < \epsilon_1u^2_2-\epsilon_2u^2_1>
+\frac{1}{2}  < \epsilon_1\eta_2+\epsilon_2\eta_1>
+\frac{\nu}{2} \frac{\partial^2}{\partial r^2} < \epsilon_1u_2+\epsilon_2u_1>~.
$$
We now demonstrate an identity that shows that this term vanishes.
Multiplying Burgers equation for $u_1$ by $\epsilon_2$ we have 
\begin{equation}
\partial [\epsilon_2 u_1]/\partial t_1 ~=~ \nu 
\partial^2[\epsilon_2 u_1/]\partial x_1^2  ~-~
\frac{1}{2} \partial [\epsilon_2u_1^2]/\partial x_1 ~+~ u_1 \eta_1 ~~~.
\end{equation}
Adding the analogous equation with the subscripts $1$ and $2$ interchanged to
the above equation and using Eqns. (\ref{derivatives}) yields the required
identity immediately.

\section{Derivation of Eqn. (19)}

We derive the equation satisfied by $<(\delta \vec u \cdot \delta \vec u)^2>$
where $\delta \vec u ~=~ u(\vec x ,t)-\vec u(\vec x',t') \equiv \vec u - \vec
u'$. We use as usual 
$$ \vec x ~=~ \vec R + \frac{1}{2} \vec r ~~~\mbox{and} ~~~ \vec x' ~=~ \vec R 
- \frac{1}{2} \vec r $$
and 
$$ t = T + \frac{1}{2} \tau ~~~\mbox{and} ~~~ t' = T - \frac{1}{2} \tau $$
leading to the identities
$$\partial/\partial x_i = \frac{1}{2}\partial /\partial R_i+\partial/\partial
r_i ~~~\mbox{and}~~~ \partial/\partial t = \frac{1}{2} \partial /\partial T +
\partial/\partial \tau$$
and the corresponding equations for the primed variables. Using these and the
Navier-Stokes equations for $u_i$ and $u_i'$ and adding them yields the basic
equation
\begin{eqnarray}
\label{c1}
2 \frac{\partial \delta u_i }{\partial \tau} ~&=&~ \nu[\nabla^2 u_i +
\nabla^{'2} u'_i] ~-~ (u_j+u'_j) \frac{\partial \delta u_i}{\partial r_j} ~-~
\delta u_j  \frac{\partial (u_i+u_i')}{\partial r_j} \nonumber \\
&& -2  \frac{\partial \delta p}{\partial r_i} ~+~ \eta_i+\eta_i' ~~
\end{eqnarray}
where $\delta p = p(\vec x , t) - p(\vec x', t')$. Multiplying this equation by
$\delta \vec u \cdot \delta \vec u ~ \delta u_i$, summing over $i$ and taking
averages over the spatially homogeneous, temporally steady state of fully
developed turbulence leads to the equation we wish to establish after
straightforward manipulations. Terms that are proportional to $\nu$ need
careful consideration. It is most convenient to extract 
the terms that are finite as $\nu
\rightarrow 0$ by using Leibniz's rule:
\begin{eqnarray}
<\delta \vec u \cdot \delta \vec u ~ \delta u_i \nabla^2 \delta u_i> ~&=&~
\partial_j < \delta \vec u \cdot \delta \vec u ~ \delta u_i\partial_j \delta
u_i> \nonumber \\
&& -<\delta \vec u \cdot \delta \vec u ~ \partial_j \delta u_i 
\partial_j \delta u_i> - 2<\delta u_i \partial_j \delta u_i \delta u_k 
\partial_j \delta u_k>
\end{eqnarray}
and similar equation for the primed variable. 
 The following identity that can be derived easily from 
incompressibility and Navier-Stokes
equation 
\begin{equation}
\nabla^2 p ~=~ -~ \frac{\partial u_i}{\partial x_j}~\frac{\partial u_j}
{\partial x_i}
\end{equation}
and the obvious result that follows from it 
$$\epsilon ~\equiv~ \frac{\nu}{2} (\partial_i u_j+\partial_j u_i)^2 ~=~ \nu
(\partial_i u_j)^2 -\nu \nabla^2 p$$
are useful in rewriting the viscous terms.


\begin{thebibliography}{99}
\bibitem{burgnu}
The rate of energy dissipation is determined by $<\eta(x)u(x)>$ which can be
evaluated using the correlations of the noise term in $k$- space to obtain
$\epsilon ~\sim~ \frac{D_0}{L} \delta^{-1-\beta}$
where $\delta$ is an inner cutoff scale (ultraviolet) 
provided physically by the
shock thickness.  The latter may be obtained by balancing the dissipative term
$\nu k^2<\hat u(k) \hat u(-k)>$ with the energy input due to the noise to
obtain 
$\nu
\propto [\frac{D_0}{L}]^{1/3}
\delta^{1-\beta/3} ~~~.$
\bibitem{burgmu2}We do not have predictions for the
(multifractal)
dynamical exponents as $\beta \rightarrow -1$; However, it is possible to argue
that the dynamical exponent for $S_4$ (in the $\tau \rightarrow 0$ limit) is
given by 
$z_4 ~=~ 1 +\zeta_4 -\zeta_5$ 
Assuming that we can equate $\tilde{z}_4 =z_4$ we obtain
$\mu_2 = 2+\zeta_4-2\zeta_5$
and $\mu_1 > \mu_2$ by the Schwarz-like inequality
$\zeta_5 ~>~ \frac{1}{2}[\zeta_4+\zeta_6]$. 
Thus $\mu=2-\zeta_6$  remains the
dominant intermittency exponent.
\bibitem{fn2}By the
von-Karman-Howarth relation one has $S_3 \propto |r|^{-\beta}$ yielding the
scaling value for $\zeta_3$. There is very good numerical support in the case 
of 
$S_2$ that $\zeta_2$ has the scaling value $-2\beta/3$; the result  is also
supported by heuristic arguments.\cite{hj1}
\end{thebibliography}
\end{document}